\documentclass[%
 aip,
 jcp,%
 amsmath,amssymb,
 preprint,%
]{revtex4-1}

\usepackage{threeparttable,array,booktabs,calc}
\usepackage{graphicx}
\usepackage{dcolumn}
\usepackage{bm}

\usepackage[version=4]{mhchem}

\usepackage{hyperref}

\usepackage{epstopdf} 

\begin{document}

\preprint{PenningPaper-1}

\title[Reactive He*-Li collisions]{Penning collisions between supersonically expanded metastable helium atoms and laser-cooled lithium atoms}

\author{Jonas Grzesiak}
\affiliation{Institute of Physics, University of Freiburg, Hermann-Herder-Str. 3, 79104 Freiburg, Germany}
\author{Takamasa Momose}
\affiliation{Department of Chemistry, The University of British Columbia, Vancouver, British Columbia V6T 1Z1, Canada}
\author{Frank Stienkemeier}
\affiliation{Institute of Physics, University of Freiburg, Hermann-Herder-Str. 3, 79104 Freiburg, Germany}
\author{Marcel Mudrich}
\affiliation{Department of Physics and Astronomy, Aarhus University, Ny Munkegade 120, 8000 Aarhus C, Denmark}
\author{Katrin Dulitz}
\email{katrin.dulitz@physik.uni-freiburg.de.}
\affiliation{Institute of Physics, University of Freiburg, Hermann-Herder-Str. 3, 79104 Freiburg, Germany}

\date{\today}

\begin{abstract}
We describe an experimental setup comprised of a discharge source for supersonic beams of metastable helium atoms and a magneto-optical trap (MOT) for ultracold lithium atoms that makes it possible to study Penning ionization and associative ionization processes at high ion count rates. The cationic reaction products are analyzed using a novel ion detection scheme which allows for mass selection, a high ion extraction efficiency and a good collision-energy resolution. The influence of elastic He-Li collisions on the steady-state Li atom number in the MOT is described, and the collision data are used to estimate the excitation efficiency of the discharge source. We also show that Penning collisions can be directly used to probe the temperature of the Li cloud without the need for an additional time-resolved absorption or fluorescence detection system.
\end{abstract}

\pacs{Valid PACS appear here}
\keywords{magneto-optical trap, MOT, laser cooling, supersonic expansion, Penning ionisation, associative ionisation, reactive scattering, cold molecules}

\maketitle

\section{\label{sec:introduction} Introduction}
Reactive collisions between electronically excited helium atoms in the 2$^3$S$_1$ and 2$^1$S$_0$ states (He*) and lithium atoms lead to the formation of \ce{Li+} ions (Penning ionization, PI, Eq.\,(1)) and \ce{HeLi+} ions (associative ionization, AI, Eq.\,(2)):
\begin{align}
\ce{He}^{*} + \ce{Li} \ce{->} [\ce{HeLi}]^{*} &\ce{->} \ce{He} + \ce{Li+} + \ce{e-}\\
                                              &\ce{->} \ce{HeLi+} + \ce{e-}
\end{align}
These reactions belong to a whole class of chemical processes in which an atomic or molecular species is ionized by a long-lived (metastable), electronically excited atom or molecule, usually referred to as Penning ionization. Such Penning processes are prevalent in diverse temperature and density regimes, where metastable species can be produced efficiently, e.g. by electron impact or charge transfer, such as in the Earth's upper atmosphere \cite{Zipf1969,Takayanagi1970,Wayne2000}, in the atmospheres of planets and their satellites \cite{Wayne2000} and in combustion and plasma processes \cite{Herron2001}. Since the chemistry of the early Universe is restricted to gas-phase reactions among the three lightest elements -- \ce{H}, \ce{He} and \ce{Li} -- the study of reactive collisions between \ce{He} and \ce{Li} is of particular relevance for an improved understanding of the formation of our Universe \cite{Lepp2002}.

Despite the important role of Penning ionization processes in Nature, the quantum-chemical treatment of Penning processes is still very difficult owing to the need for optical potentials to describe the coupling to the ionization continuum. More accurate experimental data for this reaction system will thus aid the theoretical description of potential energy surfaces and the associated quantum dynamics calculations. With a total of only five electrons involved, He*-\ce{Li} represents a collision system especially suitable for accurate quantum-chemical calculations \cite{Kedziera2015,Movre2000}. 

Penning reactions have been subject to numerous experimental investigations for more than 50 years already. The majority of these studies was done at high energies using traditional crossed-beam methods, including angle- and energy-resolved measurements of ions and electrons produced in the reaction \cite{Siska1993}. Recently, quantum resonance effects at low collision energies have been observed in merged-beam Penning collision experiments, in which the trajectories of two supersonic beams were superimposed using electromagnetic guiding fields \cite{Henson2012,Osterwalder2015}. A number of Penning ionization studies in the barrierless, ultracold s-wave scattering regime have been triggered by the quest for reaching degenerate quantum gases of metastable atoms \cite{Vassen2012}. In such experimental setups, Penning ionization leads to unwanted trap loss. It was found that Penning reactions can be efficiently suppressed by polarizing the electron spins of the trapped atoms \cite{Herschbach2000,Byron2010,Flores2016, Spoden2005}, so that even the Bose-Einstein condensation of metastable species could be achieved \cite{Robert2001,PereiraDosSantos2001}. 

The He*-\ce{Li} system has previously been studied at collision energies above 100 K using crossed- and merged-beam scattering \cite{Wang1987,Ruf1987,Merz1989}. The rate constants for He*-\ce{Li} Penning ionization were found to be very large \mbox{($\approx 10^{-9}$ cm$^3$/s)} and almost collision-energy independent \cite{Wang1987}. In addition to that, spin-statistical effects seem to strongly influence the chemical reactivity, as Penning ionization rates for He(2$^1$S$_0$)-Li(2$^2$S$_{1/2}$) scattering are 3-4 times larger than for He(2$^3$S$_1$)-Li(2$^2$S$_{1/2}$) collisions \cite{Ruf1987}. Moreover, rotational energy transfer of the He*Li quasi-molecule to the emitted electrons plays an important role, as shown in a study in which the angular dependence of Penning electron kinetic energies was determined \cite{Merz1989}.

Our approach to studying reactive He*-\ce{Li} collisions offers several benefits compared to other methods using crossed/merged beams \cite{Wang1987,Ruf1987,Merz1989,Henson2012,Osterwalder2015} and two-species traps \cite{Sawyer2008a,Sawyer2011,Parazzoli2011}. In our setup, lithium atoms are laser-cooled and confined in a magneto-optical trap (MOT) which offers the advantage of a high-density ($\approx$ 10$^{11}$ cm$^{-3}$), ultracold ($\approx$ 1 mK) and stationary scattering target, whose properties can be easily monitored and controlled using external fields. In combination with a supersonic beam source, in which typical densities of 10$^{13}$ cm$^{-3}$ are observed for the metastable atomic and molecular species, high signal count rates are observed. Moreover, the collision energy can be tuned over a wide range and the quantum states and the motion of the collision partners can be independently manipulated at spatially different positions. This latter feature can, for example, be used to probe effects of alignment and orientation on the reaction process \cite{Weiner1999}.
%

In this paper, we provide a detailed description of our experimental setup for Penning reaction studies (Section \ref{sec:experimental}). We also present a novel product ion detection scheme using a train of high-voltage pulses, and we compare this method to conventional single-pulse and continuous ion detection schemes (Section \ref{sec:multipulse}). Besides that, the decrease of the steady-state Li atom number in the MOT by elastic collisions is discussed (Section \ref{sec:elastcoll}). Using these experimental data, an estimate of the excitation ratio of the helium discharge source is obtained. In Section \ref{sec:MOTdynamics}, we demonstrate that Penning ionization can also be used for a direct temperature determination of the ultracold Li atomic cloud.

\section{\label{sec:experimental} Experimental}
\subsection{\label{sec:generalsetup} General Setup}
\begin{figure}[!htbp]
\includegraphics{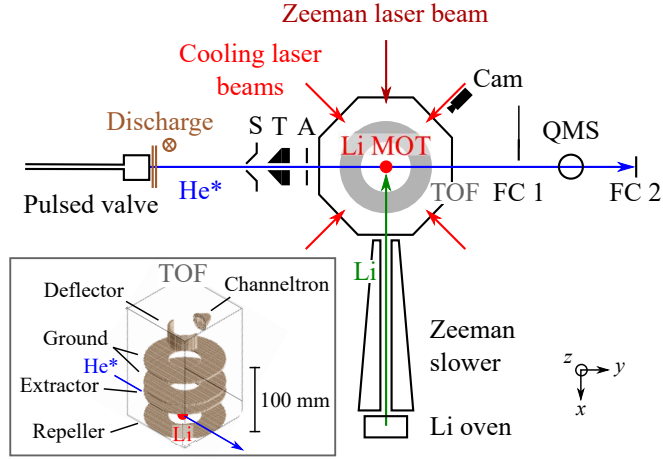}
\caption{\label{fig:sketchchamber} Sketch of the experimental setup, not to scale. Abbreviations: S = skimmer, T = torpedo-shaped aperture, A = aperture, Cam = CCD camera, TOF = ion time-of-flight detector, QMS = quadrupole mass spectrometer, FC 1 and FC 2 = Faraday cups 1 and 2. FC 1 can be moved in and out of the He* beam axis. The axes of the metastable helium beam and the lithium beam are indicated as blue and green arrows, respectively. The cooling and Zeeman laser beams are given as red- and dark-red-colored arrows, respectively. A to-scale, close-up view of the ion time-of-flight detector is shown as an inset.}
\end{figure}

The experimental setup comprises a MOT for ultracold $^7$Li atoms and a supersonic beam source for metastable helium atoms. A schematic drawing of the apparatus is shown in Figure \ref{fig:sketchchamber}. Some parts of the setup have already been described elsewhere \cite{Strebel2010,Strebel2012,Strebel2013}.

\subsubsection{\label{sec:vacuumchambers} Vacuum Chambers}
To maintain ultra-high vacuum conditions in the interaction region, the setup consists of several, differentially pumped vacuum chambers. A He source chamber, typically held at a pressure of 2$\cdot10^{-6}$ mbar during the experiment, is pumped by an oil diffusion pump (Balzers, DIF 400, 8000 l/s) and backed by a rotary vane pump (Leybold, Trivac D65 B) and a roots pump (Leybold, Ruvac WA501). For differential pumping, an intermediate chamber is used which is separated from the He source chamber and the MOT chamber by a 1-mm diameter skimmer and a torpedo-shaped, 35 mm-long aperture (1.95 $\pm$ 0.05 mm hole diameter). High vacuum of $\leq$ 1$\cdot10^{-7}$ mbar inside this chamber is attained by a \mbox{1000 l/s} turbomolecular pump (Leybold, TurboVac 1000). A home-built gate valve is used to separate the intermediate chamber from the MOT chamber. High vacuum is maintained in this small volume between the torpedo and the entrance to the MOT chamber (4-mm diameter aperture) using a 67 l/s turbomolecular pump (Pfeiffer, HP 80). The torpedo-shaped aperture, in particular, ensures a sufficient differential pumping between the intermediate chamber and the MOT chamber. A conical shape was chosen to prevent a disturbance of the helium beam. When the gate valve is opened, the pressure in the MOT chamber rises by \mbox{$<1\cdot10^{-11}$ mbar} in the presence of the helium beam. The MOT chamber, in which Penning collisions are measured, is evacuated to a base pressure of \mbox{$\leq$ 2$\cdot10^{-10}$ mbar} using two turbomolecular pumps (Leybold, TurboVac 150, 150 l/s, and Leybold, TurboVac 340M, 340 l/s). The atomic lithium source is contained within a chamber which is maintained at 2$\cdot10^{-8}$ mbar using an ion getter pump (Varian, VacIon N121T, 130 l/s).

\subsubsection{\label{sec:heliumsource} Production and Detection of Metastable Helium Atoms}
\begin{figure}[!htbp]
	\includegraphics{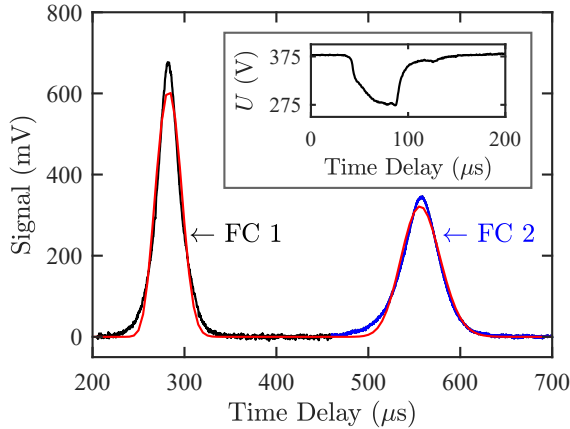}
	\caption{\label{fig:FCdet} He* beam TOF distributions as a function of time delay with respect to the valve trigger measured at the two Faraday cup detectors FC 1 (black colour) and FC 2 (blue color) and the corresponding fit curves (red color). The TOF trace consists of two consecutive measurements in which FC 1 is moved in and out of the He* beam axis, respectively. The valve temperature was 300 K. The inset shows the time dependence of the plate voltage $U$ obtained using a voltage probe. The time scale is referenced to the beginning of the valve trigger pulse.}
\end{figure}

A beam of atomic helium is produced by a supersonic gas expansion from a high-pressure reservoir (30 bar) into the vacuum using a short-pulse, high-intensity CRUCS valve (100 $\mu$m orifice diameter, 40$^{\circ}$ cone) \cite{Grzesiak2018}. He atoms inside the beam are electronically excited to the metastable $2^3$S$_1$ and $2^1$S$_0$ states using an electron-seeded plate discharge source attached to the valve exit. The discharge setup is conceptually similar to the design described by Lewandowski et al. \cite{Lewandowski2004}. It consists of two stainless steel electrode plates (1.5 mm thickness, tapered inner diameter, 4.5 mm smallest inner diameter) which are insulated by two PTFE spacers. A reliable operation of the discharge in the glow regime, i.e., without arcing, and high metastable signal intensities are achieved by setting the electrode plate at the valve orifice (1.5 mm distance to valve front plate) to 480 V, by grounding the second plate and by using an additional glow filament (OSRAM Halostar 20 W filament, 1.45 A, biased to \mbox{-100 V}). The latter leads to the emission of seed electrons which aids the ignition of the discharge \cite{Halfmann2000, Lewandowski2004}. The exact discharge settings are strongly dependent on the experimental conditions; the values given here are for guidance only. The discharge duration is limited to the effective valve opening time of $\approx$ 50 $\mu$s, which eliminates the need for an additional pulsing of the plate voltage. To ensure a stable operation of the discharge in the glow regime, the plate voltage is continuously monitored using a voltage probe (PMK, PHV 641-L). While a smooth decrease of the voltage signal is characteristic of a glow discharge (see inset to Figure \ref{fig:FCdet}), arcing is indicated by a more rapid and much higher voltage drop. To maximize He* production, we typically operate the discharge at plate voltages just below the transition to the arc regime. Since there is a sudden transition between the two discharge regimes at a certain plate voltage, a small voltage decrease (by $\approx$ 3 \%) is typically sufficient to return from the arc regime to the glow regime.

The supersonic beam consists of a $\geq$ 25 \% mixture of He(2$^1$S$_0$) and He(2$^3$S$_1$), as evidenced using optical quenching of the 2$^1$S$_0$ state in a similar experimental setup in our laboratory (not shown). This finding is in close agreement with results from other discharge sources, where singlet-to-triplet ratios on the order of 1:3 were observed \cite{Ferkel1991}.

The skimmer is placed at a distance of 10 cm from the valve opening to avoid interference effects. It is biased to a voltage of 200 V in order to deflect He atoms in Rydberg states and He ions, produced during the discharge process. Using liquid-nitrogen cooling and PID regulation (LakeShore Model 325), the valve temperature is tuned between \mbox{150.0 K} -- \mbox{300.0 K} and stabilized to better than 0.1 K. Lower temperatures could not be achieved owing to the heat load induced by the glow filament and by the oil diffusion pump.

A quadrupole mass spectrometer (QMS, Pfeiffer, QMS 200) and two stainless-steel Faraday cup plates (plate distance of 502$\pm$2 mm; amplified by a transimpedance amplifier, Femto, DLPCA-200, 10$^7$ V/A gain, 400 kHz) are located behind the collision region (Figure \ref{fig:sketchchamber}). The QMS is used to monitor the characteristics of the ground-state He beam. The Faraday cup plates are used for the accurate determination of the velocity distribution of the He* beam \cite{Christen2011}. For this, the measured He* time-of-flight (TOF) distributions at the two Faraday cup detectors are fit to a convoluted Maxwell-Boltzmann distribution \cite{Haberland1985} in a global least-squares manner. The functional form of the TOF distribution reads
\begin{equation}
f(t) = \int_{1/(t-t_0)}^{1/(t-t_1)}{\left(\frac{L}{t-t'}\right)^3\frac{A}{\sqrt{2 \pi \sigma^2}}e^{\frac{-[L/(t-t')-v_\mathrm{m}]^2}{2\sigma^2}}\mathrm{d}t'},
\label{eq:FCfiteq}
\end{equation}
where $A$ is the signal amplitude, $v_\mathrm{m}$ is the mean beam velocity and $\sigma$ is the standard deviation. The distance between the valve opening and a Faraday cup detector is given by $L$. To account for the finite opening time of the valve, the initial velocity distribution at the valve orifice is convoluted with a 30 $\mu$s long, rectangular pulse profile estimated from the width of the voltage drop signal (inset to Figure \ref{fig:FCdet}). Moreover, the beginning and the end of the discharge excitation pulse (denoted by the times $t_0$ and $t_1$ in Equation \ref{eq:FCfiteq}, respectively) are also obtained from the voltage profile.

The fitted He* beam TOF distributions are in good agreement with the experimental results (Figure \ref{fig:FCdet}). The deviations can be ascribed to the finite, irregular shape of the discharge excitation pulse or to technical artifacts, e.g., due to a bouncing of the valve plunger. Using the method described above, we are able to deduce the mean beam velocities $v_\mathrm{m}$ and the full-width-half-maximum (FWHM) velocity widths $\Delta v$ to within an uncertainty of about 0.5 \% and 10 \%, respectively. The uncertainty of the mean velocity values could be further reduced by a more accurate measurement of the discharge duration, e.g. using a photodiode, and by a more precise determination of the relative Faraday cup distance, respectively. Calculations in which the mean velocities $v_\mathrm{m}$ are obtained by dividing the relative Faraday cup distance by the time difference of the measured intensity maxima at the two Faraday cup plates yield the same results as the method described above (within 5 m/s).

\begin{figure}[!htbp]
	\includegraphics{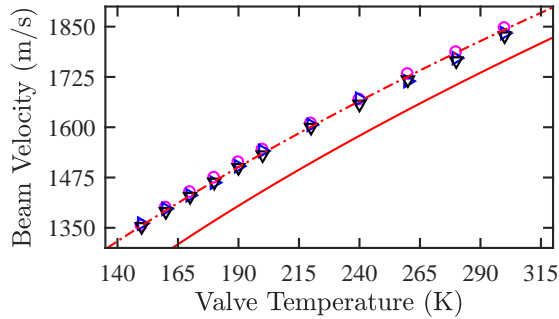}
	\caption{\label{fig:tempdep} Measured He* beam velocities $v_{\mathrm{m}}$ on three different days as a function of valve temperature (markers). The solid and dashed curves show the temperature dependence of He* beam velocity according to Equation \ref{eq:veloBeamTemp} using $T_{\mathrm{offs}}$ = 0 K and 27 K, respectively.}
\end{figure}

The He* beam velocities obtained from the solution to Equation \ref{eq:FCfiteq} range from about 1360 m/s at 150 K to \mbox{1840 m/s} at 300 K (Figure \ref{fig:tempdep}). The measured values can be matched with the expected final beam velocity \cite{Christen2008} 
\begin{equation}
v_{\mathrm{m}}=\sqrt{\frac{2 k_{\mathrm{B}}(T+T_{\mathrm{offs}})}{m} \frac{\kappa}{(\kappa-1)}},
\label{eq:veloBeamTemp}
\end{equation}
if an effective temperature offset $T_{\mathrm{offs}} = 27$ K from the monitored valve temperature $T$ is assumed. In Equation \ref{eq:veloBeamTemp}, $m$ is the helium mass, $\kappa$ = 5/3 is the helium heat capacity ratio and $k_{\mathrm{B}}$ is the Boltzmann constant. We attribute the observed large temperature offset and the broad velocity width of the He* beam ($\Delta \approx$ 110 m/s) to a significant heating of the supersonic beam by the discharge process.

We also use Faraday cup detection to quantify the flux of metastable helium atoms. While the secondary electron emission coefficients $\gamma_{\mathrm{el}}$ for He(2$^1$S$_0$) and He(2$^3$S$_1$) are very similar \cite{Dunning1971, Dunning1975a}, $\gamma_{\mathrm{el}}$ is strongly dependent on the surface finish and on the cleanliness of the surface ($\gamma_{\mathrm{el}}$ = 0.5-1.0) \cite{Dunning1971}. We have noticed different detection efficiencies on our two Faraday cup detectors, i.e., the integrated signal intensity on FC 2 is 50 \% lower than for FC 1 (Figure \ref{fig:FCdet}). This effect could, in part, be due to a partial blocking of the He* beam on FC 2 by the QMS. On the other hand, different surface properties of the two detectors cannot be ruled out. For the present experiment, we therefore assume a value of $\gamma_{\mathrm{el}} = (0.75 \pm 0.25)$ for FC 1.

\subsubsection{\label{sec:atomdetection} Production and Detection of Ultracold Lithium Atoms}
A high-density cloud of ultracold $^7$Li atoms (\mbox{$\approx 10^9$ atoms}, 3.8 mm FWHM diameter) is continuously produced using laser cooling and magneto-optical trapping. For this, an effusive beam of Li atoms is produced in an oven  ($T$ = 693 K) and decelerated in a decreasing-field Zeeman slower\footnote{Penning collisions of He* with Li atoms from the effusive beam does not result in a measurable background \ce{Li+} signal.}. A 10 mm-diameter aperture is placed in between the two chambers for differential pumping. A standard MOT configuration is used for spatial confinement of the Li atoms. The setup consists of three perpendicular, retro-reflected, circularly polarized cooling laser beams (aligned through the center of the chamber) and two coils in anti-Helmholtz configuration (wrapped around the chamber).

Laser light for the pump ($2^2S_{1/2}(F=2)$ -- $2^2P_{3/2}(F=3)$) and repump ($2^2S_{1/2}(F=1)$ -- $2^2P_{3/2}(F=2)$) laser beams  at 671 nm (saturation parameter $s_0 \approx 40$) is produced in a master-slave diode laser configuration \cite{Schuenemann1998}. The frequency of the master laser is locked to an external Li cell using saturated absorption spectroscopy. For most of the experiments described here, the MOT and the Zeeman laser frequencies are red-detuned from resonance by -50 MHz and -60 MHz, respectively.

The atomic number distribution within the cloud is inferred from the Li fluorescence intensity at 671 nm on a CCD camera (AVT Guppy F-038B NIR).

A laser beam resonant with the ($2^2S_{1/2}(F=2)$ -- $2^2P_{3/2}(F=3)$) transition (1 mm diameter, $\approx$ 100 $\mu$W) is sent through the MOT chamber to align the relative position of the Li cloud with respect to the He* beam. An elongation of the atomic cloud and a decrease of the Li fluorescence signal owing to the radiative force of this laser beam, which is sent in a direction anti-collinear with the He* beam, is a clear indicator for a good alignment between the two collision partners. Small misalignments of the Li cloud with respect to the He* beam are corrected by adjusting the current through the MOT coils and through a compensation coil on the opposite side of the Zeeman slower, respectively.

\subsubsection{\label{sec:iTOFdetector} Ion Time-of-Flight Detection of Reaction Products}
To monitor ions produced by Penning collisions, an ion time-of-flight (TOF) mass spectrometer has been built around the MOT target (see inset to Figure \ref{fig:sketchchamber}). To allow for laser beam access into the MOT chamber, the standard electrode configuration (repeller - extractor - ground) has been extended by another electrode which is used to deflect the ions onto an off-axis channeltron detector (CEM, Dr. Sjuts, Standard CEM). Owing to this non-standard electrode geometry, the electrode and CEM voltages were initially scanned over a wide parameter range to find optimum signal conditions and to avoid a saturation of the CEM at high ion count rates. Because of the magnetic-field sensitivity of the CEM, the signal intensity was also decreased when the current applied to the MOT coils was increased (Section \ref{sec:atomdetection}). To avoid this artifact in the future, we are planning to replace the CEM by an MCP detector which is much less susceptible to magnetic fields.

The ion TOF spectrometer is used in two different modes of operation: a configuration in which time-independent high voltages are applied to the repeller, the extractor and the deflector plates (DC mode) and another configuration in which the voltages to these electrodes are simultaneously toggled between high voltage and ground (pulsed mode) using three independent high-voltage switches (Behlke, HTS\textunderscore 41\textunderscore 06\textunderscore GSM).
Both modes of operation are useful for certain types of experiments. For example, the DC mode allows for a straightforward determination of the ion arrival times with respect to the valve trigger pulse, and it enables the counting of single ions. When ion counting is not necessary, we use a transimpedance amplifier (Femto, DLPCA-200, gain of 10$^5$ V/A, 500 kHz) to increase the signal-to-noise ratio for DC ion detection. The pulsed mode of operation makes it possible to discriminate between different ion masses and thus allows for a nearly background-free detection of reaction products. This is particularly relevant for our setup, since, under certain experimental conditions, we do not only observe \ce{Li+} and \ce{HeLi+} (Eqs.\,(1) and (2)), but also \ce{He+} and \ce{H2O+} ions, which are generated by intra-beam He*+He* Penning ionization and by the He*+\ce{H2O} reaction, respectively (not shown).

\section{\label{sec:results} Results and Discussion}
\subsection{\label{sec:multipulse} Multiple-Pulse Ion Detection}
In the following, a novel ion detection scheme is presented which simultaneously allows for a background-free discrimination between different ion masses, the measurement of higher ion signal intensities than using single-pulse detection and a good collision energy resolution. In this scheme, we use a train of high-voltage pulses for ion detection rather than a single pulse (Section \ref{sec:iTOFdetector}). We are only aware of one previous study in which a two-pulse ion detection sequence -- a conceptionally similar approach compared to the method described here -- was used to improve the collision-energy resolution in a scattering experiment \cite{Allmendinger2016}. A detailed description of the scheme and its extension to several detection pulses are presented here.

\begin{figure}[!htbp]
	\includegraphics{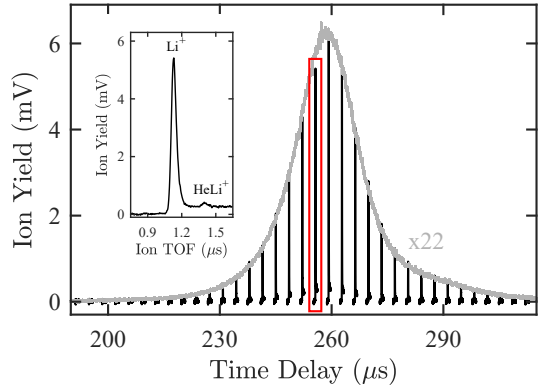}
	\caption{\label{fig:multipulse} Ion yield as a function of time delay with respect to the valve trigger measured using a train of 80 ion detection pulses (black traces) and DC ion detection (gray trace). Each detection pulse has a width of 1.65 $\mu$s, and the time gap between the pulses is set to 1.9 $\mu$s. Large amplitude noise from the high-frequency switching process is omitted for clarity. The DC ion yield is multiplied by a factor of 22 to illustrate that both the DC ion yield and the ion yield from multiple-pulse detection follow the TOF distribution of the He* beam. The inset shows a zoom into a single ion TOF profile (marked with a red box in the main figure). Here, the ion flight time is referenced to the beginning of the detection pulse. The non-zero signal offset at arrival times $> 1.2 \mu$s is due to ions produced during the detection pulse.}
\end{figure}

As a result of a multiple-pulse ion detection sequence, a mass-resolved ion signal is obtained after each high-voltage pulse, as shown in Figure \ref{fig:multipulse}. As can be seen from the inset to the figure, the \ce{Li+} ion peak from He*-Li Penning ionization (PI) has a much higher signal intensity compared to \ce{HeLi+} produced by associative ionization (AI). The AI/PI ratio is on the order of 2 \%, which is in good agreement with previous experimental studies, where a ratio of $\approx$ 0.5 \% was found at this collision energy \cite{Wang1987}. The \ce{Li+} yield from multiple-pulse detection and the DC ion yield in Figure \ref{fig:multipulse} are proportional to the number of He* atoms in the Li interaction zone, i.e., they both follow the TOF distribution of the He* beam (cf. Faraday cup signal in Figure \ref{fig:FCdet}).

A total number of 80 detection pulses is used so that the entire He*-Li interaction time is covered. The pulse width (1.65 $\mu$s) is longer than the \ce{Li+} and \ce{HeLi+} flight times to avoid that the falling edges of the high-voltage pulses disturb the ion detection. At the same time, the pulse width is kept as short as possible, since ions produced during the detection pulse duration do not contribute to the mass-resolved ion signal. Instead, they lead to a non-zero signal offset on the ion-TOF trace (see inset to Figure \ref{fig:multipulse}).

\begin{figure}[!htbp]
	\includegraphics{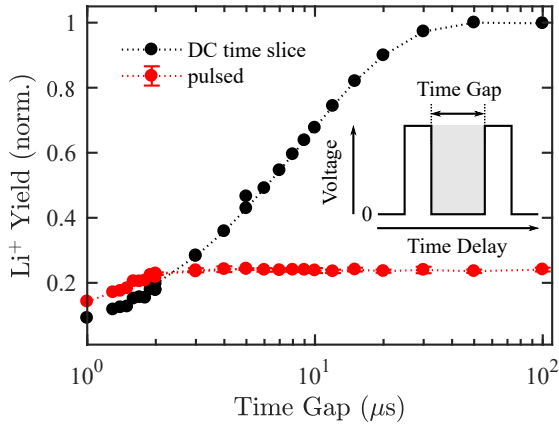} 
	\caption{\label{fig:gapvar} Integrated \ce{Li+} yield obtained from the second pulse of a two-pulse detection sequence as a function of time gap in between pulses (red markers). The first pulse is used to sweep out all ions which have accumulated in the reaction zone, so that the second pulse selectively probes the number of \ce{Li+} ions produced during the time gap in between pulses (gray-shaded region in the inset). The time delay of the second pulse with respect to the valve trigger was set to 255 $\mu$s. For comparison, the black markers show the \ce{Li+} yield from DC ion detection obtained by signal integration over the same time period as the time gap for the two-pulse sequence. Ion signal obtained outside this detection window was discarded in the post-processing of the data.}
\end{figure}

We were expecting to see an increase in the measured ion signal when the time gap between two detection pulses was increased, since \ce{Li+} ions produced during this time interval should accumulate in the detection volume. Such a behavior can be inferred from the DC ion yield, if -- during the post-processing of the data -- the DC \ce{Li+} yield is integrated over a time period corresponding to the time gap in between two detection pulses. The results obtained from this analysis (black markers in Figure \ref{fig:gapvar}) show that the integrated \ce{Li+} yield should increase up to a time gap of 30 $\mu$s. However, using pulsed ion detection (red markers in Figure \ref{fig:gapvar}), we observe that the \ce{Li+} yield only increases up to a time gap of 1.9 $\mu$s in between detection pulses. At larger time gaps, the ion yield remains nearly constant. This behavior is independent from the applied pulse duration, and a similar dependence is also observed for the \ce{HeLi+} signal (not shown). We attribute this effect to the kinetic energy release by the reaction, which leads to a rapid escape of the ions from the interaction zone. We have also observed that the bipolar high-voltage switches generate a negative bias voltage when a pulse train of more than two pulses is used ($\leq$ 1 \% of the applied high voltage). This also leads to the deflection of ions and it causes a signal decrease by another factor of two (not shown). In the future, unipolar switches will be used to eliminate this effect.

The results suggest that the maximum overall ion yield is obtained for a pulse train with a time gap of 1.9 $\mu$s and a pulse width of 1.65 $\mu$s. This results in a maximum detection efficiency of $\approx$ 50 \% compared to the integrated DC ion yield. In contrast to that, a single-pulse detection scheme always results in a low detection efficiency, since it can only extract $\approx$ 6 \% of the ions at maximum. This comparison clearly illustrates the power of the multiple-pulse technique in terms of mass-resolved ion collection efficiency.

Owing to the low temperature of the Li target (see Section \ref{sec:MOTdynamics}), the collision energy resolution in our setup is dominated by the velocity distribution of the He* beam. For the 80-pulse sequence used to obtain the traces in Figure \ref{fig:multipulse}, the sampled velocity range is determined by the time gap in between detection pulses. Hence, the effective velocity width is $\Delta v_\mathrm{eff} = \Delta v t_\mathrm{gap}/t_{\mathrm{FWHM}} = 10$ m/s, where $t_{\mathrm{FWHM}}$ is the FWHM of the ion signal intensity in DC detection mode and $t_\mathrm{gap}$ is the time gap between the detection pulses. This corresponds to an energy resolution of $\Delta E_{\mathrm{coll}}$ = 0.5 meV for a valve temperature of \mbox{300 K}. To obtain a much higher energy resolution, a smaller time gap can be used. The minimum time gap is ultimately limited by the achievable signal-to-noise ratio.

\subsection{\label{sec:elastcoll} Influence of Elastic Collisions}
\begin{figure}[!htbp]
	\includegraphics{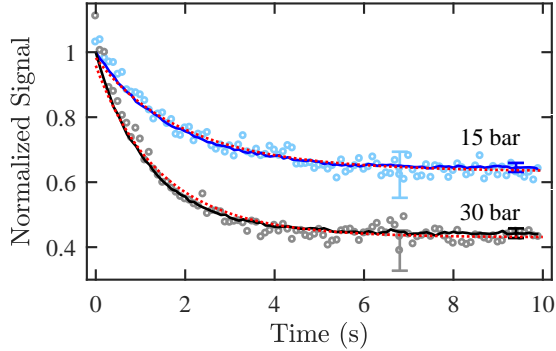}
	\caption{\label{fig:elastcollfluoresc} Time evolution of the normalized Li fluorescence signal (solid blue and black curves) and ion yield (light blue and gray markers) measured for two different He stagnation pressures at a valve repetition rate of 10 Hz. Results obtained from a least-squares fit to the fluorescence data using Equation \ref{eq:MOTDiffEq} are shown as red dashed curves. The pulsed valve is turned on at time zero. Each datapoint is an average of 15 fluorescence and ion signal determinations, respectively. The ion traces are scaled with respect to the fluorescence traces.}
\end{figure}

Many collision experiments using magneto-optical traps rely on the measurement of trap loss, in which the contribution of elastic and reactive scattering to the overall loss rate cannot be well distinguished. Our rate determinations are not influenced by elastic collision processes, since the reaction products are directly measured via ion-time-of-flight mass spectrometry (Section \ref{sec:iTOFdetector}). However, elastic He-Li collisions lead to a reduction of the steady-state Li atom number in the MOT, as observed as an initial decrease of Li fluorescence signal when the pulsed valve is turned on (solid blue and black curves in Figure \ref{fig:elastcollfluoresc}). A similar decay behavior is observed for the \ce{Li+} yield (light blue and gray markers in Figure \ref{fig:elastcollfluoresc}) which is directly proportional to the Li fluorescence signal. In the scattering experiment, we discard the ion signals of the first 50 valve shots to eliminate time-dependent signal contributions.

In our experiment, the evolution of the number of Li atoms $N_{\mathrm{Li}}$ in the MOT can be described by the differential equation
\begin{equation}\label{eq:MOTDiffEq}
\frac{dN_{\mathrm{Li}}}{dt}=L-\gamma N_{\mathrm{Li}}-\beta N_{\mathrm{Li}}^2 - \kappa'_{\mathrm{eff}} N_{\mathrm{Li}},
\end{equation}
where $L$ is the loading rate of the MOT, $\gamma N_{\mathrm{Li}}$ is the scattering rate with background gas in the chamber, $\beta N_{\mathrm{Li}}^2$ is the loss rate for Li-Li collisions inside the MOT (radiative escape and fine-structure-changing collisions, see e.g., Ritchie et al. \cite{Ritchie1995}) and $\kappa'_{\mathrm{eff}} N_{\mathrm{Li}}$ is the elastic He-Li collision rate. Since the number of metastable He atoms, $N_{\mathrm{He^*}}$, is much less than the number of ground-state helium atoms, $N_{\mathrm{He}}$, the elastic He*-Li collision rate is negligible.

The equilibrium number of Li atoms can then be expressed as
\begin{equation}\label{eq:Lieq}
N_{\mathrm{Li,eq}} = -\frac{\gamma+\kappa'_{\mathrm{eff}}}{2\beta}+\sqrt{\left(\frac{\gamma+\kappa'_{\mathrm{eff}}}{2\beta}\right)^2+\frac{L}{\beta}}.
\end{equation}
Since a pulsed He* beam is used in the experiment, $\kappa_{\mathrm{eff}} = \kappa/(f t_{\mathrm{FWHM}})$, where $f$ is the repetition rate of the pulsed valve and $t_{\mathrm{FWHM}}$ is the He* pulse width in the interaction zone obtained from the FWHM of the ion signal intensity in DC detection mode ($t_{\mathrm{FWHM}}$ = \mbox{20 $\mu$s}). To calculate $\kappa'_{\mathrm{eff}}$, the value of $\kappa_{\mathrm{eff}}$, obtained at a He stagnation pressure of 30 bar and at a 10 Hz repetition rate, is scaled by $f/(10\,\mathrm{Hz})$. The rate coefficient $\kappa = n_{\mathrm{He}}v_{\mathrm{rel}}\sigma_{\mathrm{HeLi}}$ depends on the density of the ground-state He beam $n_{\mathrm{He}}$, the relative velocity of the scattering partners $v_{\mathrm{rel}} \approx v_{\mathrm{m}} = (1832 \pm 9)$ m/s (for a He* beam at 300 K) and on the cross section for elastic scattering $\sigma_{\mathrm{HeLi}} = (127.5 \pm 3.5)\text{\AA}^2$ \cite{Dehmer1972}. The density dependence thus also explains the lower equilibrium number of Li atoms in the MOT and the lower \ce{Li+} yield at higher He valve stagnation pressures (cf. Figure \ref{fig:elastcollfluoresc}). This suggests that, at a repetition rate of 10 Hz, a He backing pressure of 15 bar is preferable for collision experiments compared to 30 bar despite a 15 \% increase in He* flux at higher stagnation pressures.

\begin{figure}[!htbp]
	\includegraphics{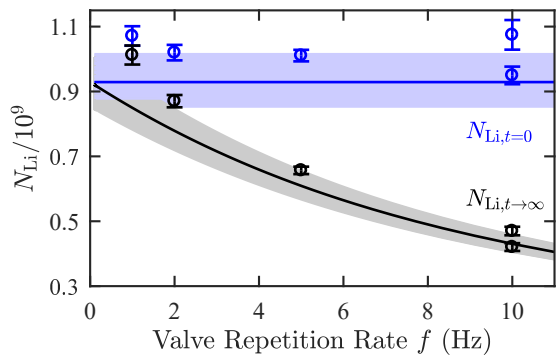}
	\caption{\label{fig:elastcollreprate} Initial and equilibrium number of lithium atoms obtained from fluorescence measurements at different valve repetition rates (blue and black markers). The solid blue (black) curve is the calculated equilibrium number of Li atoms in the absence (presence) of the He beam (30 bar stagnation pressure) obtained using Equation \ref{eq:Lieq} and the rate coefficients given in the main text. In the absence of the He beam, $\kappa = 0$. The light blue and gray shadings show the uncertainties of the solid blue and black curves, respectively, which are obtained from the uncertainties of $L$, $\beta$ and $\gamma$.}
\end{figure}

Since $\kappa_{\mathrm{eff}}$ depends on the valve repetition rate, we also observe a decrease of the equilibrium Li atom number when the valve repetition rate is increased (Figure \ref{fig:elastcollreprate}). It is advantageous for us to work at a 10 Hz repetition rate rather than, for example, at 2 Hz. At 10 Hz, the acquisition time is decreased by a factor of five, while the Li atom number is only decreased by a factor of two compared to a repetition rate of 2 Hz.

At a valve stagnation pressure of 30 bar and $f$ = \mbox{10 Hz}, we measured $\kappa'_{\mathrm{eff}} = \kappa_{\mathrm{eff}} = (0.37\pm 0.04)$ s$^{-1}$. The other rate coefficients in Equation \ref{eq:MOTDiffEq} are obtained in the absence of the He beam from the measurement of MOT loading and decay curves as $L = (2.22 \pm 0.05)\cdot10^{8}$ atoms s$^{-1}$, $\gamma=(0.07\pm 0.02)$ s$^{-1}$ and $\beta$=\mbox{$(1.83\pm0.13)\cdot 10^{-10}$ atoms$^{-1}$ s$^{-1}$}.

The time evolution of the fluorescence signal can -- in combination with the integrated Faraday cup signal -- be used as a sensitive means to determine the excitation ratio $N_{\mathrm{He^*}}/N_{\mathrm{He}}$. The number of metastable He atoms per pulse, $N_{\mathrm{He^*}}$, is obtained from the amplified and integrated signal on Faraday cup 1, $I_{\mathrm{FC1}}$ (in Vs), as
\begin{equation}
N_{\mathrm{He^*}} = \frac{I_{\mathrm{FC1}}}{\gamma_{\mathrm{el}} e G},
\label{eq:NHestar}
\end{equation}
where $\gamma_{\mathrm{el}}$ is the secondary electron emission coefficient (see above), $e$ is the elementary charge (in As) and $G$ is the gain of the signal amplifier (in V/A). The number of ground-state helium atoms per pulse in the interaction zone, $N_{\mathrm{He}}$, can then be calculated using
\begin{equation}
N_{\mathrm{He}} = n_{\mathrm{He}} v_{\mathrm{rel}} t_{\mathrm{FWHM}} A_{\mathrm{He}} = \frac{\kappa t_{\mathrm{FWHM}} A_{\mathrm{He}}}{\sigma_{\mathrm{HeLi}}}.
\label{eq:NHe}
\end{equation}
Here, $A_{\mathrm{He}} = \pi (d/2)^2$ is the cross section of the He beam in the interaction zone. The diameter of the He beam in the center of the MOT chamber, $d = 2.9\,\pm\,0.1$ mm, is determined by the skimmer diameter and by the apertures in the intermediate chamber, respectively (Section \ref{sec:vacuumchambers}). The elastic He-Li collision rate $\kappa$ is obtained by fitting the solution of Equation \ref{eq:MOTDiffEq} to the evolution of the Li atom number (cf. Figure \ref{fig:elastcollfluoresc}).

\begin{figure}[!htbp]
	\includegraphics{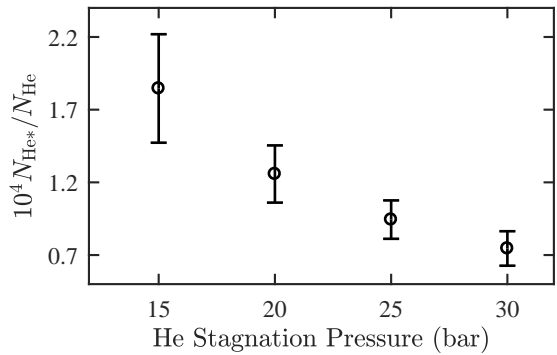}
	\caption{\label{fig:excitationratios} Excitation ratio $N_{\mathrm{He^*}}/N_{\mathrm{He}}$ determined at different He valve stagnation pressures (see main text for details). The given error bars do not include the uncertainty related to the electron emission coefficient.}
\end{figure}
Excitation ratios $N_{\mathrm{He^*}}/N_{\mathrm{He}}$ at different He valve stagnation pressures obtained using the above procedure are shown in Figure \ref{fig:excitationratios}. The excitation ratio decreases from $2\cdot10^{-4}$ at 15 bar to $7\cdot10^{-5}$ at 30 bar, which is presumably due to more efficient collisional de-excitation of He* during the supersonic expansion at higher stagnation pressures. Measurements, in which the pressure increase in the MOT chamber was recorded using a cold cathode pressure gauge, yield results for $N_{\mathrm{He}}$ that agree (within a factor of two) with the results derived from Equation \ref{eq:NHe}. Our results for the excitation ratio are a factor of three lower compared to an earlier determination by Luria et. al. \cite{Luria2009}, where $N_{\mathrm{He^*}}/N_{\mathrm{He}}\approx 6\cdot10^{-4}$ was found for a dielectric-barrier-discharge (DBD) source. We have also tested such a DBD source for He* production, but in our case, the observed He* signal intensity was more than one order of magnitude lower than the intensity obtained using a plate discharge source.

\subsection{\label{sec:MOTdynamics} Measurement of MOT Temperature using Penning Collisions}
Penning collisions can also be used as a diagnostic tool for probing the atom number and the atomic density in a MOT and in a BEC, as previously shown in several experiments \cite{Vassen2012}. Here, we add the determination of MOT temperature to this toolbox. Since temperature is one of the most important characteristics of a laser-cooled atomic sample, there exist a number of experimental methods for its measurement\cite{Adams1997}. The most common techniques probe the thermal expansion (TOF schemes)\cite{Lett1988,Weiss1989,GueryOdelin1998,Brzozowski2002,Schneble2003,Vorozcovs2005} or the velocity distribution of the ultracold cloud (release-recapture schemes)\cite{Chu1985}.

Our approach is an extension of the TOF technique. Here, the spatial overlap between an ultracold cloud and a supersonic beam of metastable atoms is monitored via Penning collisions at various time delays after the cancellation of the trapping potentials. To block the laser beams for laser cooling and Zeeman slowing, we have introduced four rapidly closing shutter heads (SRS, SR475, 40 $\mu$s closing time). The shutters, which are typically closed for a time duration of $\approx$ 7 ms, are synchronized to the trigger of the He valve. It is thus possible to trace the ballistic expansion of the ultracold Li cloud by changing the relative time delay between the triggers to the valve and to the laser shutters.

\begin{figure}[!htbp]
	\includegraphics{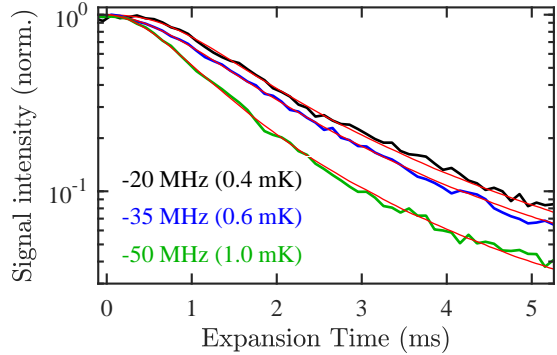}
	\caption{\label{fig:MOTexp3D} Measured integrated \ce{Li+} yield as a function of expansion time for cooling laser detunings of -20 MHz (black curve), -35 MHz (blue curve) and -50 MHz (green curve). Simulated, best-fit decay curves (see main text for details) are shown in red color. Time zero is defined as the peak arrival time of the He* beam in the Li atom volume. The cooling laser detunings and the corresponding temperatures of the Li cloud are given in the legend to the figure.}
\end{figure}
Typical decay curves of the integrated \ce{Li+} yield as a function of expansion time are shown in Figure \ref{fig:MOTexp3D} at different cooling laser detunings. The \ce{Li+} yield was obtained in a mass-selected manner by summation of all the \ce{Li+} signals in a train of 20 ion detection pulses (Section \ref{sec:multipulse}).

To model the ballistic expansion of the Li cloud, the atomic motion is numerically integrated using MATLAB. For this, three-dimensional, isotropic Maxwell-Boltzmann probability distributions are assumed for the initial spatial and velocity vectors of the Li atoms. The FWHM spatial extent of the MOT is obtained by fluorescence imaging in the presence of the Zeeman and cooling laser beams (see Section \ref{sec:atomdetection}). The temperature of the cloud $T_{\mathrm{Li}}$ is contained in the standard deviation of the velocity distribution $\sigma_v = \sqrt{k_{\mathrm{B}}T_{\mathrm{Li}}/m_{{\mathrm{Li}}}}$, where $m_{\mathrm{Li}}$ is the Li atomic mass \cite{Schneble2003}. The overlap with the He* beam is taken into account by assuming that the He* atoms are isotropically distributed within a cylindrical volume of diameter $d$ = \mbox{2.9 mm} and infinite length. A transverse offset $\Delta r_{\mathrm{offs}}$ is introduced to account for a mismatch between the center of the Li cloud and the central position of the He* beam in the $xz$ plane. Except for the temperature $T_{\mathrm{Li}}$, $\Delta r_{\mathrm{offs}}$ is the only variable parameter in the simulation. 

Decay curves, such as those shown in red color in Figure \ref{fig:MOTexp3D}, are simulated by solving the equations of motion (including gravity), and by determining the number of Li atoms within the interaction volume, at each expansion time. The temperature $T_{\mathrm{Li}}$ is obtained by comparing a large set of simulated decay curves with the experimental results in a least-squares manner. For the experimental data shown in Figure \ref{fig:MOTexp3D}, we find that $T_{\mathrm{Li}}$ = \mbox{0.4 mK} (\mbox{-20 MHz} detuning), \mbox{0.6 mK} (\mbox{-35 MHz} detuning) and \mbox{1.0 mK} (\mbox{-50 MHz} detuning). The obtained values and the temperature increase towards larger red detunings are qualitatively consistent with previous temperature determinations and with the increased trap depth predicted by calculations \cite{Schuenemann1998, Metcalf1999}.

The transverse offset was in the range $\Delta r_{\mathrm{offs}}$ = 0.3 - 0.9 mm which is coherent with the accuracy at which the Li cloud can be positioned relative to the He* beam. The temperature measurements were also reproduced using beams of other metastable noble gases. The uncertainty of the temperature determination is estimated as $\pm 20\,\%$, mainly due to the uncertainty in $\Delta r_{\mathrm{offs}}$, $d$ and due to the fluctuation of the Li atom number over the course of a measurement. Even though the uncertainty is much larger than for other MOT thermometry methods, it is still a useful technique in cases where a highly accurate temperature estimate -- requiring an additional time-resolved absorption or fluorescence detection system -- is not needed. In our case, for example, the collision temperature resolution is dominated by the velocity distribution of the He* beam. At a valve temperature of 300 K, the collision temperature resolution is $\Delta E_{\mathrm{coll}}/k_{\mathrm{B}}$ = 10 K under optimum conditions (Section \ref{sec:multipulse}) which is several orders of magnitude larger than the temperature of the Li atomic cloud.

\section{\label{sec:conclusions} Conclusions}
In this paper, we have presented a detailed characterization of a setup -- composed of a supersonic beam source and a MOT -- which can be used for the study of reactive collisions at tunable collision energies. Here, we have demonstrated that thermal Penning collisions of metastable He atoms with ultracold Li atoms can be studied with high sensitivity and high energy resolution. A Li-MOT is of particular interest for scattering measurements, since Li has a very low mass, which allows for the study of quantum resonance effects in atomic and molecular collisions at relatively high energies \mbox{($\leq 1\mathrm{K} \cdot k_{\mathrm{B}}$)} \cite{Strebel2012}. Besides that, the bosonic $^7$Li isotope can easily be replaced by fermionic $^6$Li to study spin statistical effects in the quantum scattering regime \cite{McNamara2007}. Very low collision energies could be reached if the primary beam was slowed down prior to scattering off the MOT target. This could, for example, be achieved using a molecular beam deceleration technique, such as Stark or Zeeman deceleration \cite{van2012}, or using laser cooling. The use of a MOT as a stationary, point-like target is also advantageous for angle-resolved and for quantum-state-controlled scattering experiments, respectively.

\begin{acknowledgments}
We thank the mechanical and electronics workshops in Freiburg for their commitment to this project. Technical assistance by Andrij Achkasov, Simon Hofs\"ass and Vivien Behrendt is gratefully acknowledged. This work is funded by the German Research Council (DFG) under projects DU1804/1-1 and GRK 2079. J.G. is thankful for additional financial support by the International Graduate Academy (IGA) of the Freiburg Research Services. K.D. acknowledges support by the Chemical Industry Fund (FCI) through a Liebig Fellowship.
\end{acknowledgments}


%

\end{document}